\documentclass[prl,twocolumn,superscriptaddress,longbibliography]{revtex4-2}
\usepackage[colorlinks=true, citecolor=blue, urlcolor=blue]{hyperref}
\usepackage{amsmath,amssymb,orcidlink}
\renewcommand{\section}[1]{{\par\it #1.---}\ignorespaces}

\begin{document}
\title{Anti-aging Squeezed Quantum Battery}
\author{Jing-Ci Yue\orcidlink{0009-0002-0934-3192}}\thanks{These two authors contributed equally.}
\affiliation{Key Laboratory of Quantum Theory and Applications of MoE, Lanzhou Center for Theoretical Physics, Gansu Provincial Research Center for Basic Disciplines of Quantum Physics, Key Laboratory of Theoretical Physics of Gansu Province, Lanzhou University, Lanzhou 730000, China}
\author{Jing-Tao Tang}\thanks{These two authors contributed equally.}
\affiliation{Key Laboratory of Quantum Theory and Applications of MoE, Lanzhou Center for Theoretical Physics, Gansu Provincial Research Center for Basic Disciplines of Quantum Physics, Key Laboratory of Theoretical Physics of Gansu Province, Lanzhou University, Lanzhou 730000, China}
\author{Rong-Fang Liu}
\affiliation{Key Laboratory of Quantum Theory and Applications of MoE, Lanzhou Center for Theoretical Physics, Gansu Provincial Research Center for Basic Disciplines of Quantum Physics, Key Laboratory of Theoretical Physics of Gansu Province, Lanzhou University, Lanzhou 730000, China}
\author{Jun-Hong An\orcidlink{0000-0002-3475-0729}}\email{anjhong@lzu.edu.cn}
\affiliation{Key Laboratory of Quantum Theory and Applications of MoE, Lanzhou Center for Theoretical Physics, Gansu Provincial Research Center for Basic Disciplines of Quantum Physics, Key Laboratory of Theoretical Physics of Gansu Province, Lanzhou University, Lanzhou 730000, China}

\begin{abstract}
Exhibiting the superiority of energy storage and conversion over its classical counterpart, quantum battery (QB) emerges as a promising platform to revolutionize traditional energy technologies. However, their practical performance is severely constrained by the ubiquitous decoherence in microscopic world, which causes rapid aging and reduced ergotropy. Here, we propose a decoherence-tolerant QB consisting of a continuous-variable bosonic mode. Acting as the quantum resource, the utilized squeezing endows the QB with unprecedented energy capacity and ergotropy. Through investigating the non-Markovian decoherence effect on the QB, we find that such squeezing enhanced performance can be persistently preserved under the influence of decoherence as long as a bound state is present in the energy spectrum of the total system formed by the QB and the environment. This decoherence-tolerance feature is even maintained under moderate finite-temperature condition. Providing a useful energy storage paradigm for designing long-lived QBs, our result sheds light on the realization of anti-aging QBs.  
\end{abstract}
\maketitle

\section{Introduction}
The growing demand for energy calls for more efficient energy storage and conversion strategies. To address this imperative issue, quantum thermodynamics offers a new perspective which leverages quantum effects to go beyond the classical thermodynamic limits and thus reshapes the general understanding of energy conversion at the microscale \cite{RevModPhys.93.035008,RevModPhys.83.771,PhysRevE.89.032115,PhysRevApplied.6.054014,10.1038/ncomms5185}. 
In the domain of quantum thermodynamics, quantum batteries (QBs) have emerged as high-efficiency energy storage and conversion devices \cite{RevModPhys.96.031001,10.1038/s42254-025-00906-5}, which exploit quantum resources including quantum entanglement, quantum coherence, and quantum squeezing. Besides various physical systems that have theoretically demonstrated their superiority in energy storage and conversion over classical batteries \cite{PhysRevLett.131.030402,PhysRevA.109.042424,PhysRevLett.118.150601,PhysRevLett.120.117702,PhysRevLett.125.236402,PhysRevLett.128.140501,PhysRevLett.122.047702,PhysRevLett.129.130602}, recent experimental advances in QBs based on spin systems \cite{PhysRevA.106.042601}, quantum dots \cite{PhysRevLett.131.260401}, organic systems \cite{doi:10.1126/sciadv.abk3160,Cruz_2022}, micro-cavity systems \cite{bhyh-53np,Hymas2026-mi}, ion-trap systems \cite{Hou2025-te}, nuclear magnetic resonance system \cite{PhysRevA.106.042601}, and superconducting systems \cite{sp5l-c6m8,Hu_2022,Zheng_2022,PhysRevA.109.062614,sp5l-c6m8} have highlighted the remarkable potential of QBs in energy utilization. These significant achievements are turning QBs into reality.  

Although substantial progress has been achieved in QBs \cite{PhysRevLett.131.030402,PhysRevA.109.042424,PhysRevLett.118.150601,PhysRevLett.120.117702,PhysRevLett.125.236402,PhysRevLett.128.140501,PhysRevLett.122.047702,PhysRevLett.129.130602,PhysRevE.87.042123,PhysRevE.99.052106,Friis2018precisionwork,PhysRevB.98.205423,PhysRevE.105.044125,PhysRevE.102.042111,Binder_2015,PhysRevA.106.042601,PhysRevLett.131.260401,doi:10.1126/sciadv.abk3160,Cruz_2022,sp5l-c6m8,j35s-xv5k,romero2026cumulativelocalitygovernscollective,romero2026impactthermaldissipativeeffects,f591-hx6x}, their practical implementation is severely hampered by decoherence. The interaction between QBs and the environment causes the energy loss and limits energy storage to short timescales, thereby degrading the performance of QBs. Besides the most best-known phenomenon called the aging of QB \cite{PhysRevA.100.043833}, self-discharging where a QB spontaneously loses stored energy in the absence of external load \cite{PhysRevE.103.042118,PhysRevE.105.054115,PhysRevE.109.054132,PhysRevA.109.062432}, and the ergotropic Mpemba effect where QBs prepared with higher ergotropy can discharge faster than those with lower ergotropy \cite{PhysRevLett.134.220402,5xrr-x2rm} have also gained widespread attention. Several approaches have been proposed to mitigate these environmental effects, including environment engineering \cite{43n6-rnj3,p93y-jflt,b5lx-j66h,msh1-b1cc}, dark states \cite{PhysRevApplied.14.024092,PhysRevE.100.032107}, Floquet engineering \cite{PhysRevA.102.060201}, decoherence-free subspace design \cite{doi:10.1021/acs.jpcc.9b06373}, and topological protection \cite{https://doi.org/10.1002/qute.202500897,PRXQuantum.5.030319,PhysRevLett.134.180401}. Although these schemes are experimentally accessible for short-time operations, most require continuous energy input. For the long‑term storage scenario, a passive ecoherence-tolerant design may be more suitable, which motivates our exploration. Further, most of these schemes treat the decoherence dynamics under the Born-Markov approximation, which provides an incomplete description \cite{Rivas_2014,RevModPhys.88.021002,LI20181,PhysRevA.102.012217,PhysRevLett.121.240401,PhysRevLett.122.140401,PhysRevLett.109.170402}. Recent studies have highlighted the potential advantages of non-Markovian effect on QBs \cite{Kamin_2020,PhysRevE.109.054132,mmy1-jb4n}, offering a new route to design QBs for long-term energy storage. Despite these benefits, further research is still needed to clarify the exact role of the non-Markovian effect in the energy storage of QBs. 

\begin{figure}[tbp]
\includegraphics[width=\columnwidth]{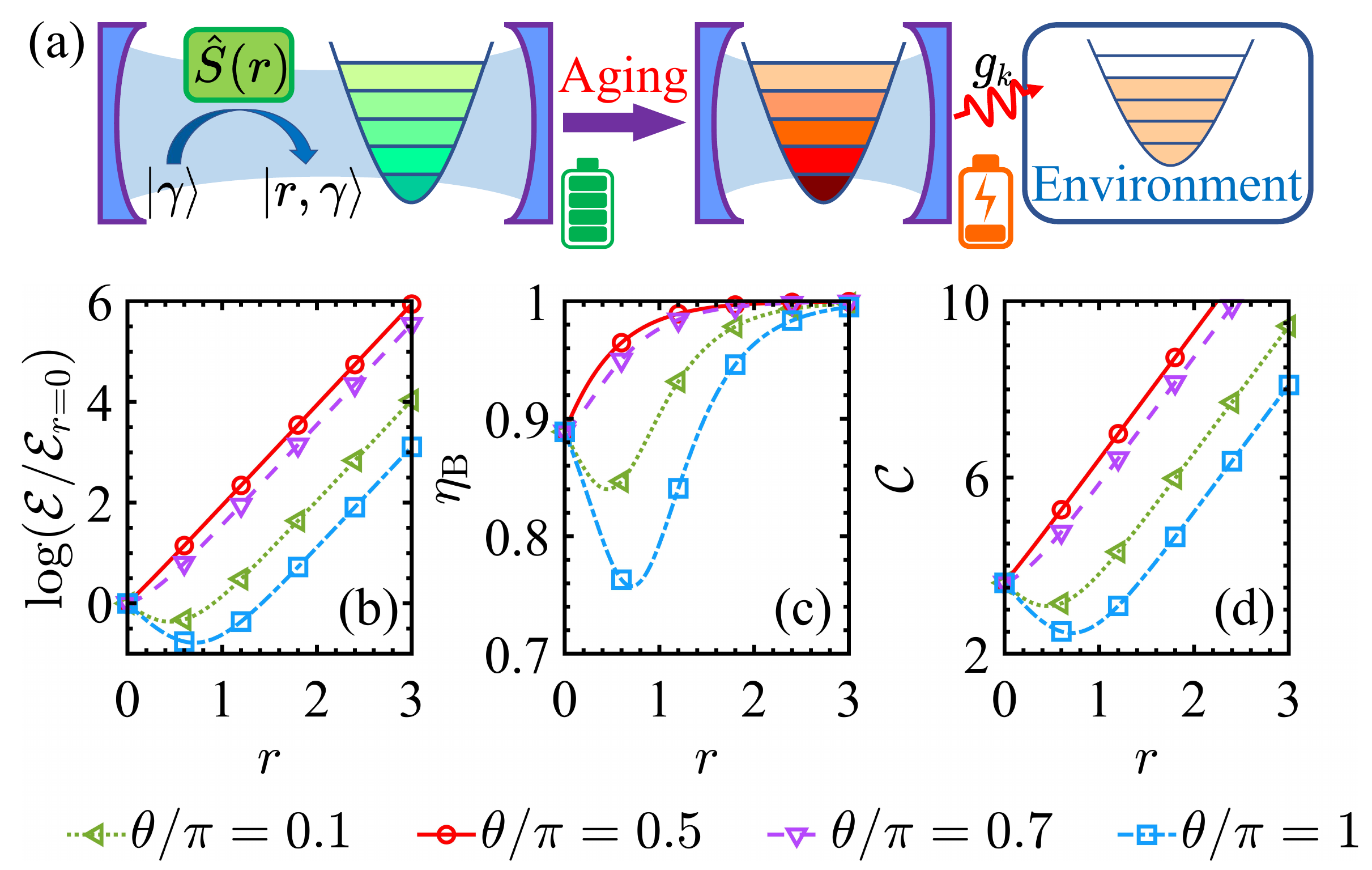}
\caption{(a) Ideal QB scheme and its aging. (b) Energy capacity, efficiency, and (c) quantum coherence  for various $\theta$ as a function of the squeezing strength $r$. We use $|\gamma|=2$.}
\label{Fig1}
\end{figure}

Inspired by research on coherence-enhanced QBs \cite{d9k1-75d4,PhysRevLett.132.090401,PhysRevE.103.042118,PhysRevLett.129.130602,s6dl-zgkx,PhysRevLett.127.100601}, we propose an energy storage paradigm for a continuous-variable QB, where a single bosonic mode initially prepared in a squeezed coherent state serves as the QB. By analyzing the long-term storage performance in the presence of decoherence, we find that the non-Markovian effect enables the QB to withstand the decoherence-induced degradation and to recover the squeezing-induced performance boost to a level close to the decoherence-free case. An examination of the energy spectrum of the total system reveals that this recovery originates from the formation of a bound state for the combined system of the bosonic mode and the environment. In contrast, when no bound state is formed, the QB evolves to a state from which no ergotropy can be extracted. Our analysis reveals that the formation of the bound state makes the quantum coherence of the QB evolves synchronously with its storage efficiency. Our results pave the way toward a long-term energy storage scheme based on the non-Markovian effect and the quantum environment engineering.   

\section{Ideal Storage Paradigm} 
We consider a continuous-variable QB realized by a single bosonic mode \cite{PhysRevLett.134.220402}. Its Hamiltonian reads $\hat{H}_B =\hbar\omega_0 (\hat{a}^\dagger \hat{a}+1/2)$ with $\hbar=1$ [see Fig. \ref{Fig1} (a)]. Although the energy stored in the QB is $\mathcal{E}(t) = \text{Tr}[\hat{H}_B \rho(t)]$, the second law of thermodynamics constrains its conversion to useful work. The maximum extractable work is quantified by ergotropy $\mathcal{W}(t) = \mathcal{E}(t) - \text{Tr}[\hat{H}_B \tilde{\rho}(t)]$, where $\tilde{\rho}(t)$ is the passive counterpart of $\rho(t)$, from which no work can be extracted by unitary operations. Thus, the storage efficiency of QB is $\eta_{\mathrm{B}}(t) =\mathcal{W}(t)/\mathcal{E}(t)$. Quantum coherence has been revealed to play a key role in the performance of QB \cite{PhysRevLett.113.140401}. It is defined by $\mathcal{C}(\rho)=S(\rho_d)-S(\rho)$, where $S(\rho)=-\mathrm{Tr}(\rho\log_2\rho)$ is the Von Neumann entropy and $\rho_d$ is the dephased state of $\rho$ \cite{PhysRevLett.125.180603}. For Gaussian states, the above quantities become
\begin{eqnarray}
\mathcal{E}(t) &=& \omega_0 \Big[ \frac{\text{Tr}[\boldsymbol{\sigma}(t)]}{4} +\frac{|\mathbf{m}(t)|^2}{2} \Big],\label{eq:energy_gauss}\\
\mathcal{W}(t) &=& \mathcal{E}(t) - \omega_0 \sqrt{\det[\boldsymbol{\sigma}(t)]}\Big/2,\label{eq:ergotropy_gauss}\\
 \mathcal{C}(t)&=&h\big((\sqrt{\mathrm{det} [ \boldsymbol{\sigma}(t) ] } -1)/2 \big)-h(\bar{n}),    \label{QC}
\end{eqnarray}
where $\mathbf{m}(t) = \text{Tr}[\rho(t) \hat{\mathbf{r}}]$ is the first-order moment vector, $\boldsymbol{\sigma}(t)$, with its elements $\sigma_{ij} =\text{Tr}[\rho(t) \{\hat{r}_i - m_i, \hat{r}_j - m_j\}]$, is the covariance matrix, $\hat{\bf{r}}=[(\hat{a}^{\dagger}+\hat{a})/\sqrt{2},i(\hat{a}^{\dagger}-\hat{a})/\sqrt{2}]$ \cite{PhysRevLett.127.210601}, $\bar{n}=(\mathrm{Tr}[ \boldsymbol{\sigma}(t) ]+2|\mathbf{m}(t)|^2-2 )/4$, and $h(x)=x\log_2x-(x+1)\log_2(x+1)$. 

In the charging process, the QB is driven to a squeezed coherent state $|\psi(0)\rangle=\hat{S}(r)\hat{D}(\gamma)|0\rangle$ such that energy is stored, where $\hat{D}(\gamma) = \exp(\gamma \hat{a}^\dagger - \gamma^* \hat{a})$, $\gamma=|\gamma|e^{i\theta}$, and $\hat{S}(r) = \exp[r(\hat{a}^2 - \hat{a}^{\dagger 2})/2]$. This can be realized by an optical parametric amplifier \cite{PhysRevA.93.033816,PhysRevA.104.053706}, reservoir engineering \cite{doi:10.1126/science.1261033}, or Josephson parametric amplifier \cite{PhysRevLett.117.020502,Fesquet2024-qi}. We focus on exploring the QB performance in the storage process. In the ideal dynamics governed by $\hat{H}_{B}$, the QB evolves to $|\psi(t)\rangle=e^{-i\omega_0t/2}\hat{S}(re^{2i\omega_0t})\hat{D}(\gamma e^{-i\omega_0t})|0\rangle$, which gives the two components of first-order moment vector as $m_1(t)+im_2(t)=\sqrt{2}(\gamma\cosh r-\gamma^*\sinh r)e^{-i\omega_0t}$ and the covariance matrix as $\boldsymbol{\sigma}(t)= \cosh 2r\mathbb{I} +\sinh 2r[\sin2\omega_0t\sigma_x-\cos2\omega_0t\sigma_z]$. Thus, we obtain $\mathcal{E}=(\bar{N}+1/2)\omega_0$, $\mathcal{W}=\bar{N}\omega_0$, and $ \mathcal{C}=-h(\bar{N})$, where $\bar{N}=|\gamma|^2(\cosh2r-\cos2\theta\sinh2r)+\sinh^2r$. These quantities are time-independent, which ensures that the QB works perfectly in the storage process. The squeezing enhanced energy capacity is characterized by $\mathcal{E}/\mathcal{E}_{r=0}=\cosh 2r-[(1-1/(2|\gamma|^2+1)]\cos 2\theta\sinh 2r$. We plot $\mathcal{E}/\mathcal{E}_{r=0}$ and $\eta_{\mathrm{B}}$ as a function of $r$ for different $\theta$ in Figs. \ref{Fig1}(b) and \ref{Fig1}(c). The results indicate that the energy capacity shows several orders of magnitude more than the squeezing-free case. The efficiency are also dramatically improved by the squeezing. Their monotonic increase with $r$ for $|\mathrm{Re}(\gamma)|\le |\mathrm{Im}(\gamma)|$ confirms the continuous enhancement. In contrast, if $|\mathrm{Re}(\gamma)|>|\mathrm{Im}(\gamma)|$, $\mathcal{E}/\mathcal{E}_{r=0}$ and $\eta_B(t)$ first decreases to its minimum and then increases with $r$. This behavior is consistent with the one of quantum coherence in Fig. \ref{Fig1}(d), manifesting the governess of quantum coherence to the performance of QB. 

\section{Dissipative storage} 
The ubiquitous decoherence arising from the interaction between the quantum system and its environment \cite{RN57} causes the aging of the QB \cite{PhysRevA.100.043833}, which severely limits its energy capacity and practical performance. The environmental effects can be classified into dephasing and dissipation. Dephasing is usually negligible in cavity QED setups. This is because dephasing typically originates from elastic collisions within a dense atomic ensemble or elastic phonon scattering in solid-state systems \cite{book}, which are unavailable in our setup. The Hamiltonian of the total system consisting of the QB and a dissipative environment is    
\begin{equation}
   \hat{H}_{\mathrm{tot}}=\hat{H}_B+\sum_{k} \big[\omega_{k}\hat{b}^{\dagger}_{k} \hat{b}_{k} +g_{k} (\hat{a}^{\dagger} \hat{b}_{k}+\hat{b}^{\dagger}_{k}\hat{a})\big],
    \label{htot}
\end{equation}
where $\hat{b}_{k}$ is the annihilation operators for the $k$th mode of the environment with frequency $\omega_k$, and $g_k$ is the coupling strength. The system-environment coupling is further characterized by the spectral density $J(\omega) = \sum_{k} |g_{k}|^{2} \delta(\omega - \omega_{k})$, which generally follows the Ohmic-family form $J(\omega)=\eta\omega^s\omega_c^{1-s}e^{-\omega/\omega_c}$ in the continuous limit, where $\eta$ is the dimensionless coupling constant, $s$ is the spectral index, and $\omega_c$ is the cut-off frequency \cite{RevModPhys.59.1}. The environment is classified into sub-Ohmic with $0<s<1$, Ohmic with $s=1$, and super-Ohmic for $s>1$. It is initially in the Gibbs state $\rho_E(0)=e^{-\beta \sum_k \omega_k \hat{b}^\dagger_k \hat{b}_k} / \mathcal{Z}$, where $\mathcal{Z}=\mathrm{Tr}e^{-\beta \sum_k \omega_k \hat{b}^\dagger_k \hat{b}_k}$ and $\beta=(k_BT)^{-1}$ is the inverse temperature, with $k_B$ being the Boltzmann constant. After tracing the environment by the Feynman-Vernon influence functional method, the QB is governed by the exact master equation \cite{PhysRevA.76.042127,PhysRevE.90.022122,PhysRevApplied.17.034073} 
\begin{align}
\hspace{-0.1cm}\dot{\rho}(t) = i \Omega(t) [\rho(t),\hat{a}^\dagger \hat{a}] +\big[ (\Gamma_t + \Gamma^\beta_t) \mathcal{L}_{\hat{a}}+ \Gamma^\beta_t \mathcal{L}_{\hat{a}^\dagger}\big] \rho(t),  \label{dotrho}
\end{align}
where $\mathcal{L}_{\hat{o}}\cdot = 2 \hat{o} \cdot \hat{o}^\dagger - \{\hat{o}^\dagger \hat{o}, \cdot\}$ is the Lindblad superoperator, $\Omega(t) = -\text{Im}[\dot{u}(t)/u(t)]$ is the renormalized frequency, $\Gamma_t = -\text{Re}[\dot{u}(t)/u(t)]$ and $\Gamma^\beta_t = \dot{v}(t)/2 +  v(t)\Gamma_t$ are the dissipation and noise coefficients. The functions $u(t)$ and $v(t)$ obey
\begin{align}
\dot{u}(t) &+ i\omega_0 u(t) + \int_0^t d\tau \, \mu(t - \tau) u(\tau) = 0,  \label{ut} \\
v(t) &= \int_0^t d\tau_1 \int_0^t d\tau_2 \, u^*(\tau_1) \nu(\tau_1 - \tau_2) u(\tau_2), \label{vt}
\end{align}
where $u(0)=1$, $\mu(t)=\int_0^{\infty} d\omega J(\omega) e^{-i\omega t}$, and $\nu(t) = \int_0^{\infty} d\omega J(\omega)e^{-i\omega t}/(e^{\beta\omega}-1)$. Through its time-dependent coefficients, Eq. \eqref{dotrho} manifests the memory effect inherent in the non-Markovian dynamics generated by the convolution in Eqs. \eqref{ut} and \eqref{vt}. 

Since $\rho(t)$ remains Gaussian, its first moment $\mathbf{m}(t)$ and the covariance matrix $\boldsymbol{\sigma}(t)$ fully characterize the dynamical evolution. Solving Eq. \eqref{dotrho} gives $m_1(t)+im_2(t)=\sqrt{2}(\gamma\cosh r-\gamma^*\sinh r)u(t)$ and $\boldsymbol{\sigma}(t) = [1 + 2v(t) + 2\sinh^2 r |u(t)|^2] \mathbb{I} - \sinh 2r\{\mathrm{Im}[ u(t)^2]\sigma_x+\mathrm{Re}[ u(t)^2] \sigma_z \}$. Thus, substituting them into Eqs. \eqref{eq:energy_gauss}-\eqref{QC} gives
\begin{eqnarray}
\mathcal{E}(t) &=& [\bar{n}(t)+1/2]\omega_0, \label{Edissipation} \\
\mathcal{W}(t) &=&\mathcal{E}(t) -\vartheta(t)\omega_0/2,\label{dedd}\\
\mathcal{C}(t)&=& h\Big((\vartheta(t)-1)/2\Big)-h(\bar{n}(t)),\label{cohdd}
\end{eqnarray}
where $\bar{n}(t)=v(t)+|u(t)|^2\bar{N}$ and $\vartheta(t)=\{[1+2v(t)+2\left|u(t)\right|^2\sinh^2r]^2-\left|u(t)\right|^4\sinh^22r\}^{1/2}$.

In the weak-coupling regime, applying the Born-Markov approximation to Eq. \eqref{ut} leads to $u_{\text{MA}}(t) = \exp\left \{ -\kappa t-i\left [\,\omega_0 + \Delta(\omega_0)\,  \right ]t \right \}$ and $v_{\text{MA}}(t) = \bar{n}_0(1 - e^{-2\kappa t})$, where $\kappa=\pi J(\omega_0)$ and $\Delta(\omega_0) = \mathcal{P} \int_0^{\infty} \frac{J(\omega)}{\omega_0 - \omega} d\omega$ \cite{PhysRevE.90.022122}. Here, $\mathcal{P}$ denotes the Cauchy principal value and $\bar{n}_0=(e^{\beta\omega_0}-1)^{-1}$. $\Gamma_t$ and $\Gamma_t^\beta$ become positive constants, e.g. $\Gamma_{t,\mathrm{MA}}=\kappa$ and $\Gamma_{t,\mathrm{MA}}^\beta=\kappa\bar{n}_0$. Thus, the QB evolves exclusively to a thermal state $\rho_{\text{MA}}(\infty) = e^{-\beta \hat{H}_{\mathrm{B}}} / \text{Tr}(e^{-\beta \hat{H}_{\mathrm{B}}})$ in the long-time limit regardless of the initial state. Because the thermal state is a passive state, the ergotropy $\mathcal{W}(t)$ approaches zero even though the stored energy $\mathcal{E}_{\mathrm{MA}}(\infty)=\left [ v(\infty)+1/2 \right ] \omega_0$ is nonzero. This is further witnessed by the vanishing quantum coherence $\mathcal{C}(\infty)=0$, demonstrating that none of the energy stored by the QB can be extracted. Thus, the Markovian decoherence destroys the squeezing enhanced performance of the QB. This is called the aging of QB.    

Going beyond the Born-Markov approximation, we now evaluate the exact environmental impact on the QB. Equation \eqref{ut} is analytically unsolvable. Its long-time form is obtained by the Laplace transform $\tilde{u}(p)=\int_0^{\infty}u(t)e^{-pt}dt$, which converts Eq. \eqref{ut} to $\tilde{u}(p)=\big[p+i\omega_{0}+\int_{0}^{\infty} d\omega J(\omega)/(p+i\omega)\big]^{-1}$. $u(t)$ is derived by making the inverse Laplace transform to $\tilde{u}(p)$, which requires identifying the poles via  
\begin{equation}
    y(\tilde{E})\equiv \omega_0 - \int_{0}^{\infty} d\omega \frac{J(\omega)}{\omega - \tilde{E}} = \tilde{E}, ~~(\tilde{E}=ip). 
    \label{poles}
\end{equation}
It is interesting to find that the roots of Eq. \eqref{poles} are the eigenenergies of Eq. \eqref{htot} in the single-excitation subspace up to a constant $\omega_0/2$. This can be proven by applying the eigenstate expanded as $|\Phi\rangle=(x\hat{a}^{\dagger}+\sum_{k}y_{k}\hat{b}_{k}^{\dagger})|0,\{0_{k}\}\rangle$ to $\hat{H}_{\mathrm{tot}}|\Phi\rangle=E|\Phi\rangle$, which yields Eq. \eqref{poles} in the continuous limit with $\tilde{E}=E-\omega_0/2$. Since $y(\tilde{E})$ monotonically decreases for $\tilde{E}<0$, Eq. \eqref{poles} has one isolated root $\tilde{E}_b$ as long as $y(\tilde{E})<0$. The corresponding eigenstate of $E_b=\tilde{E_b}+\omega_0/2$ is called the bound state. For $\tilde{E}>0$, $y(\tilde{E})$ is not analytic because the integral in $y(\tilde{E})$ diverges. Thus, it has infinite roots forming a continuous energy band. Using the poles derived from Eq. \eqref{poles} and the residue theorem, we have $u(t) = Ze^{-i\tilde{E}_bt} + \int_{0}^{\infty} d\tilde{E} \Theta(\tilde{E}) e^{-i\tilde{E}t}$, where $Z = \big[ 1 + \int_{0}^{\infty} J(\omega)/(\tilde{E}_b - \omega)^2 d\omega \big]^{-1}$ and $\Theta(\omega)=J(\omega)/\big\{[\omega-\omega_0-\Delta(\omega)]^2+[\pi J(\omega)]^2\big\}$. The first term arises from the potentially formed bound state. The second term originates from the energy band and approaches zero in the long-time limit due to the out-of-phase interference. Therefore, when the bound state is absent, we have $\lim_{t\to\infty} u(t)=0$, describing a complete decoherence. When the bound state is formed, we have $\lim_{t\to\infty} u(t)=Ze^{-i\tilde{E}_bt}$, indicating the decoherence suppression. The condition of forming the bound state for the Ohmic-family spectral density is calculated from $y(0)<0$ to be $\omega_0<\eta\omega_c\underline{\Gamma}(s)$, with $\underline{\Gamma}(s)$ being the Euler’s gamma function. Moreover, $v(t)$ tends to $v(\infty)=\int_{0}^{\infty}\big(e^{\beta\omega}-1\big)^{-1}\big\{\Theta(\omega)+\big[Z^2J(\omega)/(\omega-\tilde{E}_b)^2\big]\big\}d\omega$ \cite{PhysRevLett.133.050401}. It is easy to find from Eqs. \eqref{dedd} and \eqref{cohdd} that both $\mathcal{W}(t)$ and $\mathcal{C}(t)$ converge to zero by using $\lim_{t\to\infty} u(t)=0$ when the bound state is absent, which leads to the aging of the QB as the Markovian case. Conversely, when the bound state is present, we have
\begin{eqnarray}
    \bar{n}(\infty)&=&v(\infty)+Z^2\bar{N},\label{nmk1}\\
    \vartheta(\infty)&=&\{[1+2v(\infty)+2Z^2\sinh^2r]^2-Z^4\sinh^22r\}^{1\over2}.~~~~~\label{nmk2}
\end{eqnarray}
It is remarkable to find that the formation of the bound state produces nonzero ergotropy $\mathcal{W}(\infty)$ and quantum coherence $\mathcal{C}(\infty)$, which restores the squeezing enhanced performance of the QB. This supplies an intrinsic mechanism to preserve the performance of the squeezed QB in the presence of dissipative environment.  

\begin{figure}[tbp]
\centering
\includegraphics[width=\columnwidth]{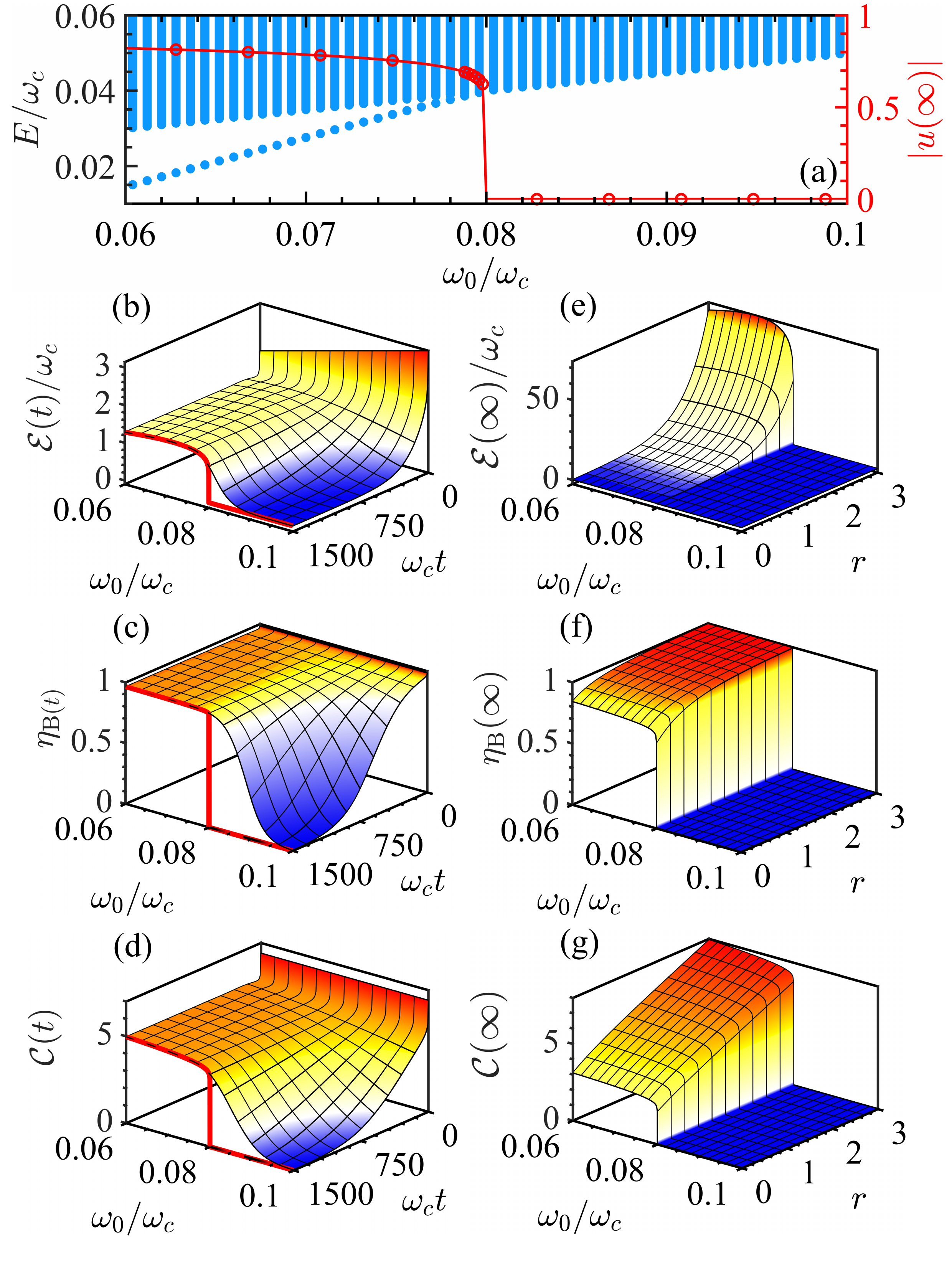}
\caption{(a) Energy spectrum obtained from Eq. \eqref{poles} (blue dots), $Z$ obtained analytically (red line), and $|u(t=1000\omega_c^{-1})|$ numerically calculated by Eq. \eqref{ut} (red circles). Evolution of (b) $\mathcal{E}(t)$, (c) $\eta_{\mathrm{B}}(t)$, and (d) $\mathcal{C}(t)$ in different $\omega_0$. The red curves are the analytical long-time behavior. (e) $\mathcal{E}(\infty)$, (f) $\eta_{\mathrm{B}}(\infty)$, and (g) $\mathcal{C}(\infty)$ as functions of $r$ and $\omega_0$. We use $\eta=0.08$, $s=1$, $\beta=0$, $r=1$, and $\gamma=2i$. }  
\label{Fig2}
\end{figure}

\begin{figure}[tbp]
\centering
\includegraphics[width=\columnwidth]{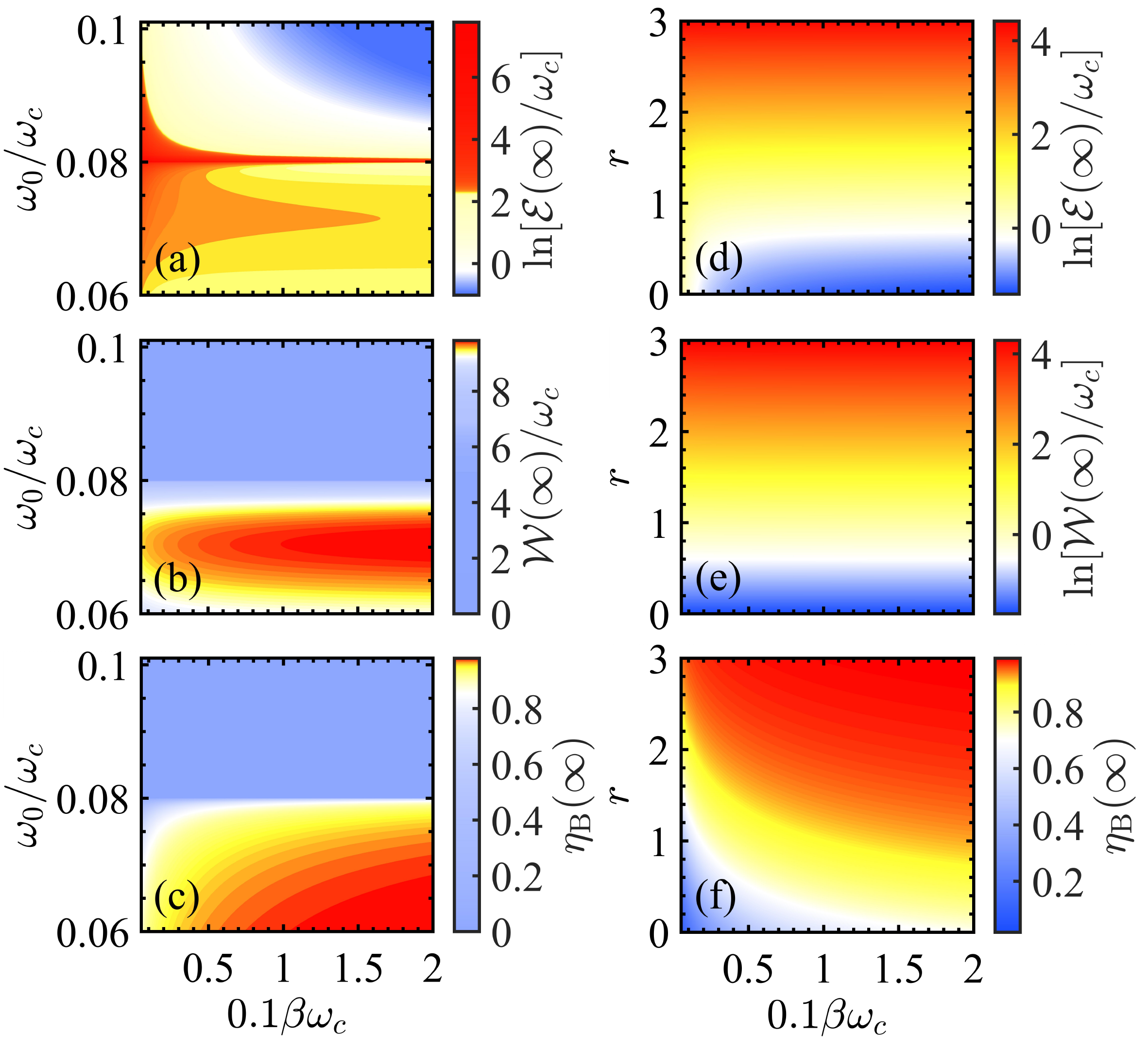}
\caption{Effect of the finite temperature on the QB. Long-term behavior of (a) the stored energy, (b) the ergotropy and (c) the storage efficiency for various $\omega_0/\omega_c$ with $r=2$. Long-time behavior of (d) the stored energy, (e) the ergotropy and (f) the storage efficiency for various $r$ with $\omega_0/\omega_c=0.07$. We use $\gamma=2i$, $\eta=0.08$ and $s=1$. }    
\label{Fig3}
\end{figure}
\section{Numerical results} To validate our anti-aging continuous-variable QB scheme, we calculate the energy spectrum of the total system by numerically solving Eq. \eqref{poles}. As shown in Fig. \ref{Fig2}(a), the bound state arises and $|u(\infty)|$ becomes nonzero when $\omega_0/\omega_c<\eta$, which corroborates our analytical result. The evolution of the energy, QB storage efficiency, and the quantum coherence in Figs. \ref{Fig2}(b), \ref{Fig2}(c), and \ref{Fig2}(d) clearly demonstrates the tolerance of our QB to the decoherence in the presence of the bound state. All of them decay to zero when the bound state is absent, whereas they are preserved at stable values when the bound state emerges. The long-time behavior of $\mathcal{W}(t)$, $\eta_{\mathrm{B}}(t)$, and $\mathcal{C}(t)$ match well with the values evaluated from Eqs. \eqref{nmk1} and \eqref{nmk2} [see the red lines in Figs. \ref{Fig2}(b), \ref{Fig2}(c), and \ref{Fig2}(d)]. It confirms the dominance of the bound state in the steady-state performance of our QB. Figures \ref{Fig2}(e) and \ref{Fig2}(f) show that the squeezing enhanced energy capacity and efficiency are recovered in the steady state when the bound state is formed. At $\theta=0.5\pi$, the energy achieves an exponential enhancement and the storage efficiency approaches $1$ as $r$ increases. Besides, increasing $r$ also improves the quantum coherence in this situation, see Fig. \ref{Fig2}(g). Therefore, the bound state makes our QB exhibit the unique anti-aging feature. On the other hand, when the bound state is absent, although a finite energy is still stored, the ergotropy and thus the efficiency are zero. 

We now extend the analysis to finite temperature. As shown in Fig. \ref{Fig3}(a), the stored energy decreases with $\beta$ and undergoes an exponential decay near $\omega_0/\omega_c=0.08$. Both the ergotropy and the QB storage efficiency increase with $\beta$ in the presence of the bound state, and $\mathcal{W}(\infty)$ reaches its maximum near $\omega_0/\omega_c=0.07$, which are shown in Figs. \ref{Fig3}(b) and \ref{Fig3}(c). Since the efficiency is determined by the ratio of ergotropy to stored energy, a reduction in $\mathcal{E}(\infty)$ together with an increasing $\mathcal{W}(\infty)$ leads to an enhanced $\eta_{\mathrm{B}}(\infty)$. However, $\mathcal{E}(\infty)$ decreases with $\omega_0/\omega_c$ while $\mathcal{W}(\infty)$ and $\eta_{\mathrm{B}}(\infty)$ remain zero for any $\beta$ when the bound state is not formed. Having established the role of the bound state, we investigate the effect of the finite temperature on the enhancement brought by squeezing in the presence of the bound state. Figures \ref{Fig3}(d)-(f) show that the effect of squeezing persists at various $\beta$ with the existence of the bound state. Specifically, Fig. \ref{Fig3}(d) shows that the increase of the stored energy with $r$ becomes weaker as $\beta$ increases, whereas the increase of ergotropy is almost unaffected [see Fig. \ref{Fig3}(e)]. Consequently, the storage efficiency increases with $r$ for different $\beta$ [see Fig. \ref{Fig3}(f)]. Therefore, although finite temperature reduces the stored energy, the ergotropy and storage efficiency do not show a significant degradation over the range of $\beta$ considered here and beneficial effect of squeezing remains present as long as the bound state exists.

\begin{figure}[tbp]
\includegraphics[width=\linewidth]{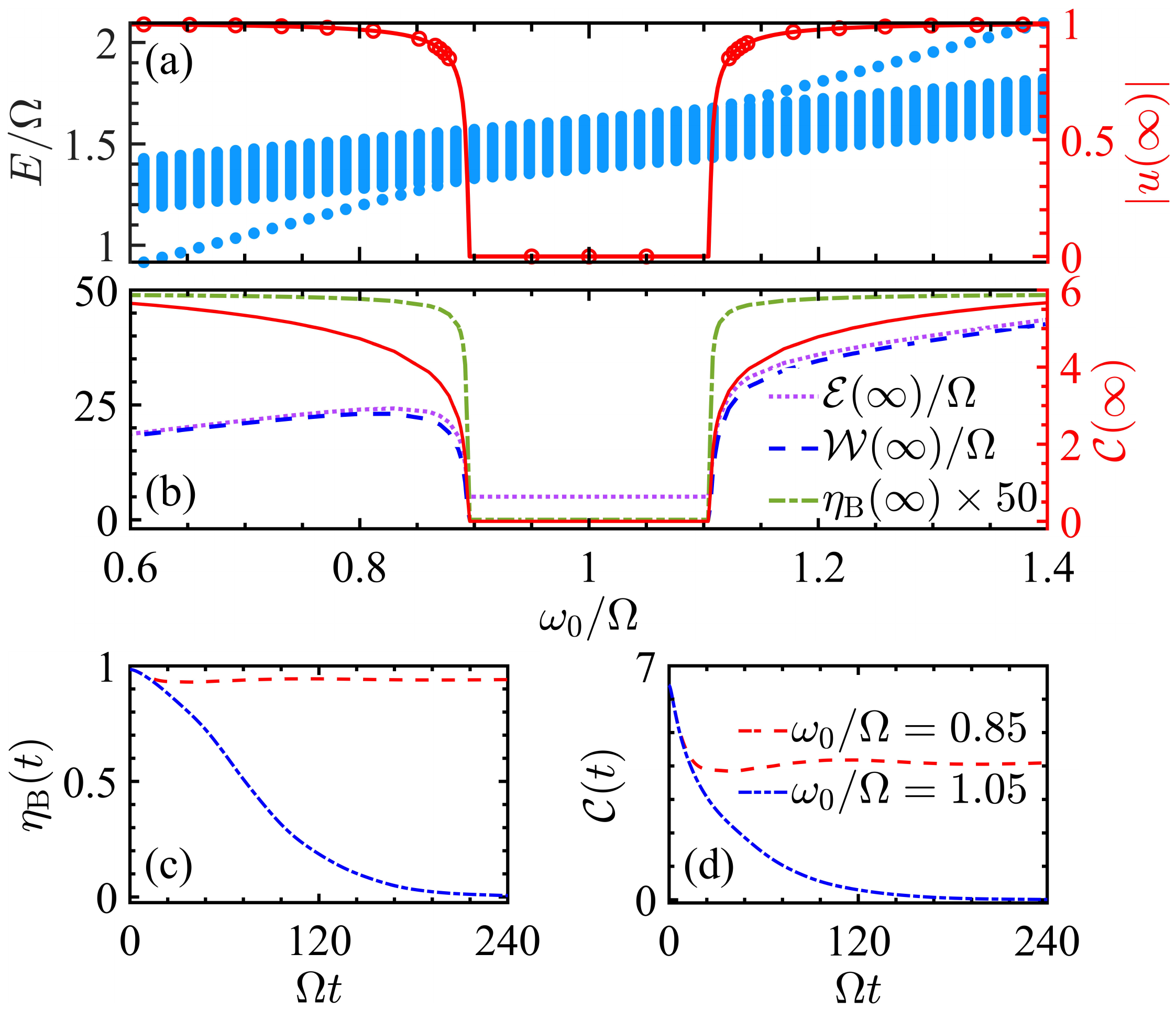}    
\caption{(a) Energy spectrum of a resonator interacting with an array of coupled resonators (blue dots), $Z$ obtained analytically (red line), and $|u( t=200\Omega^{-1})|$ numerically calculated by Eq. \eqref{ut} (red circles). (b) Long-time behaviors of $\mathcal{E}(t)$, $\mathcal{W}(t)$, $\eta_{\mathrm{B}}(t)$, and $\mathcal{C}(t)$. Evolution of (c) $\eta_{\mathrm{B}}(t)$ and (d) quantum coherence. We use $g=0.03\Omega$ and $\zeta=0.06\Omega$. } 
\label{Fig4}
\end{figure}

\section{Physical Realization}
Crucially, although our discussion assumes an Ohmic spectral density, the mechanism remains valid for other spectra since for any given spectral density there always exists a frequency range that supports the formation of the bound state. Our scheme is therefore directly applicable to state-of-the-art strong-coupling platforms, such as a resonator interacting with an array of coupled resonators that serves as the environment, which has been realized in microdisk cavities \cite{Barclay2006}. The environment and the interaction Hamiltonians are $\hat{H}_E=\sum_{j=1}^{M} \big[ \Omega \hat{b}_j^\dagger \hat{b}_j + \zeta \big( \hat{b}_j^\dagger \hat{b}_{j+1} + \hat{b}_{j+1}^\dagger \hat{b}_j \big) \big]$ and $\hat{H}_I=g\big( \hat{a}^\dagger \hat{b}_1 + \hat{a} \hat{b}_1^\dagger \big)$. Transforming the environment Hamiltonian into momentum space by the Fourier transform yields $\hat{H}_E=\sum_k\omega_k\hat{b}_k^\dagger\hat{b}_k$, with the dispersion relation $\omega_k=\Omega+2\zeta\cos k$, which has a finite bandwidth of $4\zeta$. The corresponding spectral density takes the form $J(\omega) = g^2/(2 \pi \zeta) \sqrt{4 \zeta^2 - (\omega - \Omega)^2}$. This finite-bandwidth structure provides a powerful platform for exploiting the non-Markovian effect. The numerical result in Fig. \ref{Fig4}(a) confirms that, as long as the bound state exists in the energy spectrum, $|u(\infty)|$ becomes a nonzero constant. Consequently, the energy, the ergotropy, and the storage efficiency in the steady state become nonzero, which is consistent with the behavior of the quantum coherence, see Fig. \ref{Fig4}(b). The dynamical results in Fig. \ref{Fig4}(c) verify that $\eta_{\mathrm{}B}(t)$ decays from $0.98$ to zero when the bound state fails to appear but is preserved to be $0.93$ when it emerges. The evolution of quantum coherence in Fig. \ref{Fig4}(d) qualitatively matches the results in Fig. \ref{Fig4}(c), manifesting the dominant role played by quantum coherence in the QB performance.   

 \section{Conclusion}
In summary, we have proposed a squeezing enhanced continuous-variable QB in a single bosonic mode. The QB exhibits strong tolerance to the environment induced decoherence, which endows our QB with a perfect anti-aging feature. Our analysis reveals that the ideal performance is persistently kept as long as a bound state is present in the energy spectrum of the total system formed by the bosonic mode and its environment. This mechanism enables the QB to work in a high efficiency and stabilizes the quantum coherence without degradation. Moreover, the performance of the QB remains satisfactory even in a finite-temperature environment. Refreshing our understanding of the decoherence effect on QB, our work provides a practical framework for mitigating the QB aging and facilitates the development of long-lived QBs. 

\begin{acknowledgements}
The work is supported by the National Natural Science Foundation of China (Grant No. 92576202), the Quantum Science and Technology-National Science and Technology Major Project (Grant No. 2023ZD0300904), the Natural Science Foundation of Gansu Province (Grants No. 26RCKA011 and No. 25JRRA799), the Fundamental Research Funds for the Central Universities (Grant No. lzujbky-2026-jdzx02), and the ‘111 Center’ (Grant No. B20063). 
\end{acknowledgements}

\bibliography{reference}

\end{document}